\documentclass{article}
\usepackage{graphicx}

\begin{document}

\def\be{\begin{equation}}
\def\ee#1{\label{#1}\end{equation}}

\title{Irreversible processes and the accelerated-decelerated 
phases of the Universe}
\author{G. M. Kremer\footnote{kremer@fisica.ufpr.br}$\,$ and 
M. C. N. Teixeira da  Silva\footnote{navarro@fisica.ufpr.br}\\
Departamento de F\'\i sica, Universidade Federal do Paran\'a,\\
Caixa Postal 19044, 81531-990 Curitiba, Brazil}
\date{}
\maketitle

\begin{abstract}
A model for the Universe is proposed where it is considered
as a mixture of scalar and matter  fields. The particle 
production is due to an irreversible transfer of energy from the gravitational 
field to the matter field and represented by a non-equilibrium pressure.
 This model can 
simulate three distinct periods of the Universe: (a) an accelerated epoch  
where the energy density of the scalar field prevails over the matter
field, (b) a past decelerated period 
where the energy density of the matter field becomes more
predominant than the scalar energy density,
 and (c) a present acceleration phase 
where the scalar energy density overcomes the energy density of the 
matter field.
\end{abstract}

\section{Introduction}

In order to take into account the irreversible processes during the
evolution of the Universe, cosmological models were proposed  where
the Einstein field equations are combined with the field equations
of the thermodynamic theory of 
irreversible processes. For a homogeneous and
isotropic Universe, represented by the Robertson-Walker metric,
there exists only one term related to the irreversible processes, namely
the non-equilibrium pressure. 
In Eckart (or first-order) thermodynamic theory the non-equilibrium 
pressure is considered as a constitutive quantity which is proportional 
to the Hubble parameter
and whose proportionality  factor  is the coefficient bulk viscosity
(see e.g. the works~\cite{eck1,eck2,eck3}), whereas in extended 
(second-order or causal) thermodynamic theory the
non-equilibrium pressure is supposed to obey an evolution equation 
(see e.g. the works~\cite{ext,ext1,ext2,ext3,ext4,KD0,KD1,K,KW}).
These theories are known in the literature as viscous cosmological models.

According to cosmological observations one can distinguish three 
different periods of the Universe with respect to its acceleration field, 
namely, 
(a) an early accelerated epoch, (b) a past decelerated period, and 
(c) a present accelerated phase.

The early acceleration of the Universe, which refers to the inflationary 
period, 
can be modeled by a mixture of a scalar field -- the so-called inflaton -- 
and a matter field
while the non-equilibrium pressure is identified with the particle 
production due to an irreversible transfer of energy from the gravitational 
field to the matter field~\cite{KD0,Z,Z1}. 
The inflaton is supposed to have a negative pressure
which is the responsible for the acceleration of the early Universe. 

The present acceleration of the Universe is also due to a scalar field with a 
negative pressure, since it
has been observed that the energy density of the present Universe is not due
to matter or radiation but to an extraordinary non-baryonic matter and energy.
Hence, the present Universe can be  modeled as a mixture of a matter field 
with a scalar field which is identified with the dark energy density. 
If the
irreversible processes are taken into account the non-equilibrium pressure,
in this case, is the responsible for the transfer of energy between the 
matter and gravitational 
fields. There exist at 
least two  candidates in the literature for the dark energy density, namely 
the quintessence~\cite{quin1,quin2,quin3,quin4,KD1} and the Chaplygin gas
(see e.g.~\cite{K} and the references therein). 

The objective of this work is to present a model of the Universe which could 
describe the distinct periods of the Universe, beginning with an 
early accelerated epoch, passing through a decelerated period and leading 
back to an accelerated phase. For that end we  model a 
homogeneous and isotropic Universe as a mixture of 
two constituents, namely:
(a) a scalar field that plays the role of the inflaton at the early 
accelerated period
and of the dark energy at the present accelerated phase; (b) a matter field 
that represents the particles 
classically and which are created from the irreversible transfer of energy from
the gravitational field to the matter field. 
Here we show, among other results, that this model can simulate
the three distinct periods: the early accelerated, the past decelerated 
and the present accelerated periods. Units have been chosen so that 
$c=\hbar=k=1$.

\section{Field Equations}

The equations of state that 
relate the pressures of the scalar field $p_\phi$ and  of the matter field 
$p_m$ to  their   
energy densities $\rho_\phi$ and  $\rho_m$ are given 
by
\be
\cases{
p_\phi=w_\phi\rho_\phi,\,\,\qquad\hbox{with}\qquad -1\leq w_\phi\leq 0,\cr 
p_m=w_m\rho_m,\qquad\hbox{with}\qquad0\leq w_m\leq1.}
\ee{a}
While the equation of state of the matter is the well-known 
barotropic equation with $w_m=0;1/3;2/3$ representing a pressureless fluid, 
radiation and non-relativistic matter, respectively, the  motivation for the 
equation of state of the
scalar field can be found e.g. in the works~\cite{Jus1,Jus2} 
and in the references therein.

If we consider the irreversible processes of particle production
in the  Universe, the energy-momentum tensor $T^{\mu\nu}$ is written as
\be
T^{\mu\nu}=(\rho+p+\varpi)U^\mu U^\nu-
(p+\varpi)g^{\mu\nu}.
\ee{1}
In the  above equation
$U^\mu$ (such that $U^\mu U_\mu=1$) is the four-velocity  
and  $\varpi$ denotes the non-equilibrium pressure which  is the quantity 
responsible  for particle production~\cite{KD0,Z,Z1} during the 
evolution of the Universe. Moreover, 
the pressure $p$ and the energy density $\rho$ of the mixture 
are given by 
\be
\rho=\rho_\phi+\rho_m,\qquad p=p_\phi+p_m.
\ee{2}

>From the conservation law of the energy-momentum tensor
${T^{\mu\nu}}_{;\nu}=0$ it follows the balance equation for the energy 
density of the mixture that -- in a comoving frame and by 
considering the Robertson-Walker metric -- reads
\be
\dot\rho+3H(\rho+p+\varpi)=0.
\ee{3}
The quantity $H=\dot a(t)/a(t)$ is the Hubble parameter while
$a(t)$ denotes the cosmic scale factor and the over-dot  
refers to a differentiation with respect to time.

The  cosmic scale factor is connected 
with the energy density of the mixture by the Friedmann equation
\be
H^2+{k\over a^2}={8\pi G\over3}\rho,
\ee{4}
where $G$ is the gravitational constant and $k$ assumes the values $+1,0,-1$ 
for closed, flat and open Universes, respectively.

We assume that the scalar field interacts only with itself and is 
minimally coupled to the gravitational field.
In this case, the  balance equation for the energy density of the 
scalar field  decouple from the 
energy density of the mixture (\ref{3}) and can be written as
\be
\dot\rho_\phi+3H(\rho_\phi+p_\phi)=0.
\ee{5}
Equation (\ref{5}) is used to get from (\ref{3}) the balance equation 
for the  energy density of the matter field 
\be
\dot\rho_m+3H(\rho_m+p_m)=-3H\varpi.
\ee{6}
>From the above equation it is possible to interpret the term $-3H\varpi$ 
as the energy density production rate of the matter field 
(see e.g.~\cite{KD1,K}). 

The relationship between 
the cosmic scale factor and the energy density of the scalar field
can be obtained from the integration
of (\ref{5}) by considering the equation of state given in (\ref{a}), 
yielding
\be
\rho_\phi=\rho_\phi^0\left({a_0\over a}\right)^{3(w_\phi+1)}.
\ee{8}
In  the above equation  $\rho_\phi^0$ is the value of the
energy density of the scalar field at $t=0$ (by adjusting clocks) while $a_0$ 
is the corresponding value of the cosmic scale factor.
 
Now we differentiate
the Friedmann equation  (\ref{4})  with respect to time and 
get the following equation for the time evolution  
of the cosmic scale factor 
$$
\dot H+{3\over 2}(w_m+1)\left(H^2+{k\over a^2}\right)={k\over a^2}
$$
\be
+4\pi G
\left[(w_m-w_\phi)\rho_\phi^0
\left({a_0\over a}\right)^{3(w_\phi+1)}
-\varpi\right].
\ee{9}
In order to find a solution of (\ref{9}) for the 
cosmic scale factor one has to 
specify how the non-equilibrium pressure $\varpi$ is connected with $a(t)$.
Here we assume that $\varpi$ is a variable within the framework of extended 
(causal or second-order) thermodynamic theory and write the 
linearized evolution equation for the non-equilibrium pressure as
\be
\varpi+\tau\dot\varpi=-3\eta H.
\ee{10}
For the derivation of this equation within the framework of the Boltzmann
equation one is referred to~\cite{CK}.  Moreover,
the coefficient of bulk viscosity $\eta$ and
the characteristic time $\tau$ are assumed to be related to the energy 
density of the mixture $\rho$ by 
\be
\eta=\alpha\rho,\qquad\hbox{with}\qquad\tau={\eta/ \rho},
\ee{11}
where $\alpha$ is a constant (see e.g.~\cite{ext,KD0,KD1,K}). 

For the solution of the system of differential equations (\ref{9}) 
and (\ref{10})  we introduce the 
dimensionless quantities
\be
H\equiv {H\over H_0},\quad t\equiv tH_0,\quad a\equiv {a\over a_0},\quad
\varpi\equiv {8\pi G\varpi\over 3H_0^2}, 
\ee{12}
and the dimensionless coefficients
\be
\alpha\equiv\alpha H_0,\qquad \chi=H_0^2a_0^2,
\ee{13}
where the index zero denotes the value of the quantity at $t=0$ 
(by adjusting clocks).

Now the system of differential equations (\ref{9}) and (\ref{10})
can be written in terms of the  dimensionless quantities  
(\ref{12}) and (\ref{13}) as
$$
\dot H+{3\over 2}(w_m+1)\left(H^2+{k\over \chi a^2}\right)={k\over\chi a^2}
$$
\be
+{3\over 2}
\left[(w_m-w_\phi){1+k/\chi\over 
1+\rho_m^0/\rho_\phi^0}
\left({1\over a}
\right)^{3(w_\phi+1)}-\varpi\right],
\ee{15}
\be
\varpi+\alpha\dot\varpi
=-3\alpha H\left(H^2+{k\over \chi a^2}\right).
\ee{16}

With respect to the above dimensionless quantities 
the energy densities of the scalar  and 
matter fields, can be determined from  (\ref{8}) and (\ref{4}) and read
\be
{\rho_\phi\over\rho_\phi^0}=\left({1\over a}\right)^{3(w_\phi+1)},
\ee{17}
\be
{\rho_m\over\rho_\phi^0}={1+\rho_m^0/\rho_\phi^0\over 1+k/\chi
} \left(H^2+{k\over \chi a^2}\right)-\left({1\over a}\right)^{3(w_\phi+1)}.
\ee{18}

By specifying initial conditions  for  $a(t)$, 
$\dot a(t)$ and $\varpi(t)$ at time $t=0$ and values for the 
parameters $w_\phi$, $w_m$, $\rho_m^0/\rho_\phi^0$, $k$, $\chi$
and $\alpha$ one can determine
from the system of differential equations  (\ref{15}) and (\ref{16})
the time evolution of the cosmic scale factor $a(t)$ and of 
the non-equilibrium pressure $\varpi(t)$. Once $a(t)$ is a known 
function of time the energy densities are obtained from 
(\ref{17}) and (\ref{18}).
The parameters $w_\phi$, $w_m$, $\rho_m^0/\rho_\phi^0$, $k$, $\chi$ 
and $\alpha$  have the following interpretation:
(a) $w_\phi$ and $w_m$  represent the ratios
between the pressures and the energy densities of the scalar field and matter 
fields, respectively; (b) $\rho_m^0/\rho_\phi^0$ gives the ratio between the 
initial amount of the energy density of the matter field with respect
to that of the  scalar field; 
(c) $k$ and $\chi$ are related with the space-time geometry; and (d) 
$\alpha$ is connected with the relevancy of the irreversible
processes that correspond to the particle production.

\section{Results and Discussions}

\begin{figure}
\begin{center}
\includegraphics[width=7.5cm]{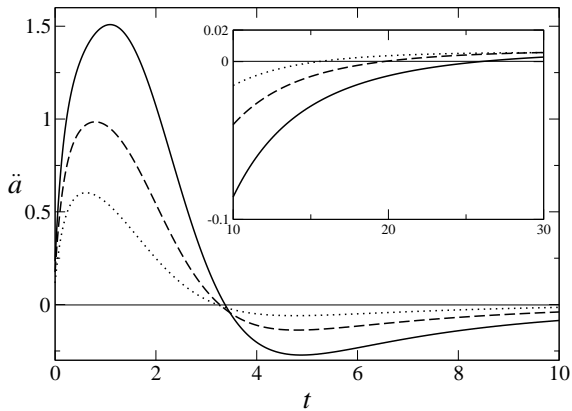}
\caption{Acceleration $\ddot a$ vs time $t$ for closed
(straight line), flat (dashed line) and open (dotted line) Universes.}
\end{center}
\end{figure}

We have solved  the system of differential equations (\ref{15}) and
(\ref{16}) numerically
by considering the following initial conditions: $a(0)=1$ for the
cosmic scale factor, $H(0)=1$ for the Hubble parameter and 
$\varpi(0)=0$ for the non-equilibrium pressure. There still remains 
much freedom to find the solution of the two systems of differential equations,
since they do depend on the parameters $w_\phi$, $w_m$, 
$\rho_m^0/\rho_\phi^0$, $k$, $\chi$ and $\alpha$. 
The figures 1 and 2 were obtained by
choosing: (a) $w_\phi=-0.45$ and  $w_m=1/3$  for the ratios
between the pressures and the energy densities of the scalar field and 
of the matter fields, respectively; (b) $\rho_m^0/\rho_\phi^0=0$, i.e., 
a vanishing initial amount of the energy density of matter field since the 
particles are created through the irreversible process of energy 
transfer from the gravitational field to the matter field;
(c) $\chi=3$ and $k=0,+1,-1$ for the importance of the space-time
geometry and (d) $\alpha=0.3$ 
for the influence of the irreversible processes in the particle
production. We call attention to the fact that $w_m=1/3$ refers to a radiation
field and the condition $w_\phi=-0.45$ satisfies the restriction 
for the quintessence $w_\phi<-1/3$ (see e.g.~\cite{Jus2}). Below 
we shall comment how the change of these parameters 
affect the solution  of the system of differential equations.

We have plotted in figure 1  the time evolution of the acceleration  field
whereas in figure 2 it is shown the time evolution of the energy densities 
for the scalar and matter fields in the cases of 
closed (straight line), flat (dashed line) and open (dotted line) 
Universes. These figures follow from the numerical solution of the 
system of differential equations  (\ref{15}) and
(\ref{16}) for the cosmic scale factor and for the non-equilibrium pressure
and from the equations (\ref{17}) and (\ref{18}) 
for the energy densities.
 One can infer from  figures  1 and 2 that 
there exist three distinct periods: (a) an accelerated epoch
where the energy density of the scalar field prevails over the matter
field, (b) a past decelerated period where the energy density of the 
matter field becomes more predominant than that of the  scalar field, 
and (c) a present accelerated phase where the scalar field energy density 
overcomes the energy density of the matter field. Hence, the transition from
the early accelerated period to the decelerated epoch is connected 
with the transition from a scalar field to a matter field dominated Universe, 
whereas the transition
from the decelerated phase to the present accelerated epoch is connected with
the transition from a matter field to a scalar field dominated Universe.
In the first accelerated period the closed Universe has the 
largest value for the acceleration field followed by the flat and the open
Universes. One can understand this behavior by recognizing from 
the equation for the non-equilibrium pressure (\ref{16}) 
that  $\varpi$  has  the largest negative value
for the closed Universe followed by the flat and the open
Universes which implies that the acceleration,  given by  
\be
{\ddot a\over a}=-{4\pi G\over 3}(\rho+3p+3\varpi),
\ee{as}
becomes  the largest also for the closed Universe.
The deceleration is the largest for the closed Universe 
since  more matter is created in a closed Universe (see figure 2). 
This behavior is a direct consequence of the fact that the 
non-equilibrium pressure is the responsible for
the creation of the matter field and it is larger in a closed Universe.
The present acceleration is attained first by the open Universe followed by 
the flat and closed Universes since the energy density of the scalar field
decays more slowly and lesser matter field is created in the case of the 
open Universe than the two others.
 
\begin{figure}
\begin{center}
\includegraphics[width=7.5cm]{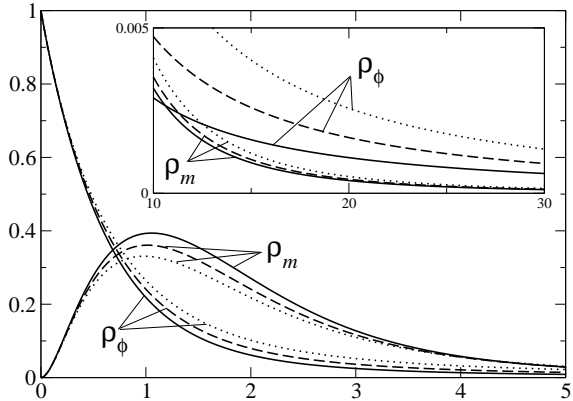}
\caption{Energy densities of scalar $\rho_\phi$ and matter $\rho_m$ fields
vs time $t$ for closed  (straight lines), flat (dashed lines) and 
open (dotted lines) Universes.}
\end{center}
\end{figure}

We have the following remarks concerning the change of the parameters: 
(a) by decreasing the values of the parameter for the scalar field $w_\phi$  
the energy density of the scalar field decays more slowly 
(see equation (\ref{17})) so that the initial accelerated period 
increases and the deceleration period decreases;
 (b) when the value of the
parameter $w_m$ for the matter field is increased in 
the interval between radiation and non-relativistic matter $1/3\leq w_m\leq2/3$
the energy density of the matter field decreases (see equation (\ref{6}))
so that there exists a smaller decelerated period; (c)
when $w_m$ is decreased in the interval between dust and radiation  
$0\leq w_m\leq1/3$ this behavior reverses;
(d) by decreasing the value of the
parameter $\chi$ -- which refers to the importance of the space-time geometry 
-- the period of past deceleration becomes larger for the closed Universe 
and smaller
for the open Universe, since for the former case the non-equilibrium pressure
becomes more negative than that of the latter (see equation (\ref{16})) ; 
(e) by increasing the
value of the parameter $\alpha$ -- which is liable  for the significance of
irreversible processes -- the 
period of past deceleration becomes larger for the closed Universe and smaller
for the  open Universe, which
 confirms that the irreversible processes are
connected with the particle production and that the energy density 
of the matter field answer
for the decelerated period. From the above remarks we infer that the 
behavior of the solutions found here for the acceleration field 
and for the scalar and matter energy densities is also valid by changing  
the values of the 
parameters $w_\phi$, $w_m$, $\rho_m^0/\rho_\phi^0$, $k$, $\chi$ 
and $\alpha$ within specified ranges.

\end{document}